\documentclass{article}
\usepackage{graphicx}
\usepackage[utf8]{inputenc}
\usepackage{hyperref}
\usepackage{booktabs}
\usepackage{listings}
\usepackage{color} 
\definecolor{mygreen}{rgb}{0,0.6,0}
\definecolor{mygray}{rgb}{0.5,0.5,0.5}
\definecolor{mymauve}{rgb}{0.58,0,0.82}
 
\usepackage[margin=1.5in]{geometry}

\title{BCD: A Cross-Architecture Binary Comparison Database Experiment Using Different Locality Sensitive Hashing Algorithms}
\author{Haoxi Tan \\ \href{mailto:haoxi.tan@uq.net.au}{haoxi.tan@uq.net.au}}

\date{October 2021}

\begin{document}

\maketitle

\begin{abstract}

Given a binary executable without source code, it is difficult to determine what each function in the binary does by reverse engineering it, and even harder without prior experience and context. In this paper, we performed a comparison of different hashing functions' effectiveness at detecting similar lifted snippets of LLVM IR code, and present the design and implementation of a framework for cross-architecture binary code similarity search database using MinHash as the chosen hashing algorithm, over SimHash, SSDEEP and TLSH. The motivation is to help reverse engineers to quickly gain context of functions in an unknown binary by comparing it against a database of known functions. The code for this project is open source and can be found at \href{https://github.com/h4sh5/bcddb}{https://github.com/h4sh5/bcddb}.

\textbf{Keywords:} reverse engineering, binary code similarity, code clone detection, minhash, llvm, cyber security, locality sensitive hashing, fuzzy hashing

\end{abstract}

\section{Introduction}

Reverse engineering is important in cyber security for purposes such as analyzing malicious code and finding vulnerabilities in software. The process of reverse engineering involves first gaining an overview of the functions in a given binary, which often requires context of what they do and prior experience \cite{observationsRE}. Although there are many mature tools in a reverse engineer's tool belt to disassemble, decompile and analyze binaries\cite{ghidra,kvroustek2017retdec,capa}, there is a gap in open source tooling that performs one-to-many comparisons in binary code against a known code base to quickly gain awareness of what each function might do.

\subsection{The Reverse Engineering Process}

The process of (software) reverse engineering is performed to determine the original purpose of a piece of software, usually in binary form without source code available. There are various tools and techniques used by reverse engineers, such as disassemblers, decompilers and debuggers to both statically and dynamically inspect the software. An observational study of different processes reverse engineers take\cite{observationsRE} revealed that they approach unfamiliar programs from a non-linear, fact-finding perspective. They make hypotheses about program functionality using existing ``beacons" (clues and patterns), then focus on proving or disproving them. 

During the aforementioned study, when asked to articulate the process they take when approaching a new program to reverse, all of the participants (N=16) responded that they start by seeking a high-level program view, and that the results from this step guides which part of the program to prioritize for more complex investigation. Having established initial questions and hypotheses based on the overview step, they then perform sub-component scanning to only review small parts of the program (e.g. functions) at a time to produce more refined hypotheses and questions, before finally moving onto in-depth static and dynamic analysis to prove their theories and answer their questions. 

The primary aim of BCD is to add tooling into the first stage of the reverse engineering process: to provide an overview of unknown functions in the binary in context of known functions in the database, to better guide the reverse engineer's efforts in the second phase (sub-component scanning). 

\subsection{Binary Code Similarity}

Compiled binary code is ISA (Instruction Set Architecture) specific, with popular architectures being Intel x86/64, ARM, MIPS etc. There has been many approaches in literature for binary code similarity, some able to detect cross-architecture similarities. An in-depth survey of this field was done by Haq and Caballero\cite{haq2021survey}. Being able to detect cross-architecture similarities has a big advantage, given that malware compiled for devices using non-mainstream architectures for Internet of Things and Operational Technology are on the rise. 

Approaches taken by prior works to perform cross-architecture binary code similarity can be summarised roughly by the following list (and often used in combination):
\begin{enumerate}
    \item Comparing CFGs (Control Flow Graphs) or ASTs (Abstract Syntax Trees) using graph edit distance or subgraph isomorphism. This approach is very popular among works targeting cross-architecture vulnerability detection\cite{feng2016scalableGENIUS,eschweiler2016discovre}.
    \item Training machine learning models based on embedding program features such as instructions, Input/Output pairs or CFGs into vectors\cite{xu2017neuralGEMINI}, and/or using a neural network (such as LTSM and Siamese networks)\cite{zuo2019neuralInnereye,xu2017neuralGEMINI};
    \item Using Locality Sensitive Hashing approaches to compare syntactic or semantic instruction features \cite{feng2016scalableGENIUS,pewny2015crossMultiMH}
    \item Overcome architecture-independence by first lifting the assembly code to an Intermediate Representation \cite{pewny2015crossMultiMH,ding2016kam1n0,chandramohan2016bingo}
\end{enumerate}

There has been an increase in cross-architecture binary similarity approaches focused around identifying the same vulnerabilities across different architectures\cite{feng2016scalableGENIUS,xu2017neuralGEMINI,zuo2019neuralInnereye}, such as firmware images, due to the rise of Internet of Things. However, there hasn't been much work that utilizes binary code similarity specifically for aiding the process of understanding binary code in a reverse engineering context; the closest is Rendevouz \cite{khoo2013rendezvous}, which specifically aimed to build a binary code search engine. Rendevouz utilized DynInst, an open source binary instrumentation library to perform code analysis, and its approach is not cross-architecture. Unfortunately, the implementation was never released to public, which are the two issues this work aims to address.

\subsection{Intermediate Representations}

Intermediate Representations (IR) are used in both the compilation and decompilation process\cite{cifuentes1994reverse}, to serve as a layer between low level architecture dependent assembly code and high-level source code to make control flow analysis and data flow analysis easier. Examples of IRs appear everywhere in compilers, static and dynamic analysis tools, such as LLVM IR, Ghidra's PCODE, Valgrind's VEX etc. 

IR is very useful for eliminating architecture dependence for code similarity comparisons, as it can both be compiled from high level source code and lifted from low level assembly code to provide a common ground for comparison.

\subsection{Locality Sensitive Hashing}

Cryptographic hashing algorithms (such as MD5 and SHA1) are used widely to compare equivalence in two pieces of data. However, they do not capture similarities in data. There are two common ``fuzzy" hashing approaches that detect similarities in data: Locality Sensitive Hashing (LSH) and Context-Triggered Piecewise Hashing (CTPH).

Locality Sensitive Hashing hashes data in same sized pieces; as in, a piece of data (such as \texttt{Hello World}) is divided into a number of tokens (\texttt{Hel, llo, loW, Wor..}) and then each token is hashed, producing a set of hashes. Then different techniques are used to reduce the dimensions of the set to a smaller set or a single number, by using pseudo-random selection (e.g. MinHash\cite{broder1997resemblanceMinHash} randomly chooses the $p$ numbers in the set using a consistent seed). 

CTPH is similar, but instead of hashing individual tokens over a rolling window, it is a rolling hash algorithm that produces a pseudo-random output based only on the current context of an input \cite{2006ssdeep}. 

Fuzzy hashing approaches have been utilized before in detecting binary similarity, such as in Multi-MH\cite{pewny2015crossMultiMH}, which used MinHash to perform similarity searching across function input/output pairs.

\section{Approach}
\label{approach} 

The general approach of BCD to construct a cross-architecture binary code similarity database is as follows:

\begin{enumerate}
    \item Take the input binary, preferably with symbols (not stripped)
    \item Lift it to LLVM IR using RetDec\cite{kvroustek2017retdec}, an open source decompiler
    \item Extract individual functions out of the LLVM IR code, and convert it into a set of instruction tokens through a series of normalization transforms
    \item Run the transformed IR tokens through a fuzzy hashing algorithm
    \item Store the hashes and functions in a database, and compare functions against each other using the hashes
    \item For faster indexed lookups, index functions using hashes as the index into a hash table and perform hash lookups to find similar functions
\end{enumerate}

A comparison of hashing functions are performed at the end of step 5 using function names as ground truth (hence why binaries with symbols are preferred). The dataset used to compare the hashing functions are samples of Mirai botnet across 9 different architectures (arm4vl, arm5vl, arm6vl, i586, i686, x86, mips, mipsel, powerpc), totaling to 893116 compared function pairs across all architectures, with 3473 having the same name. 

Four different fuzzy hashing algorithms are compared: MinHash\cite{broder1997resemblanceMinHash}, SimHash\cite{charikar2002similaritySimHash}, SSDEEP (CTPH)\cite{2006ssdeep} and TLSH\cite{oliver2013tlsh}.

\subsection{IR Normalization}
\label{irnormal}

Normalization of the output LLVM IR instructions needed to be done to aid cross-architecture similarity searching, because the LLVM IR lifting process can introduce ISA-specific differences such as different integer word sizes across 16,32 and 64 bit architectures, offset order based stack variable and global variable names (since RetDec uses Single Static Assignment\cite{van2007staticSSA} when lifting assembly to LLVM IR, assigning new variable names on each use). 

Different ISAs may have different number of stack variables due to different number of registers and stack sizes being used, leading to change in the IR code when lifted from the assembly code.

The goal of the IR Normalization phase is to capture only the important semantic information that maximizes chances that a function would be matched across a different ISA. The transformations applied to the lifted LLVM IR code are:
\begin{itemize}
    \item Single Static Assignment references and register names (e.g. \texttt{\%2}, \texttt{\%eax}) are replaced with a generic reference token \texttt{\%r} 
    \item Generated variable names with offset numbers are replaced with only the variable type (e.g. \texttt{\%stack\_var\_-4} replaced with \texttt{\%stack\_var})
    \item Integer types and sizes dropped, since data type recovery is imperfect and word sizes differ between ISAs (e.g. \texttt{i16}, \texttt{i32} are dropped)
    \item Instruction addresses (for debugging), comments and other extraneous artifacts generated by the tool are dropped
\end{itemize}

For the sake of demonstration, this snippet of \href{https://gist.github.com/rverton/a44fc8ca67ab9ec32089#file-rc4-c-L54}{RC4 encryption function}\cite{rc4gist} (which calls two other stages of RC4) from Github is compiled into three different architectures (ARM, MIPS and x86) with GCC then lifted by RetDec and fed into the normalization process.

The C code of the function is:
\begin{lstlisting}[language=C]
int RC4(char *key, char *plaintext, unsigned char *ciphertext) {

    unsigned char S[N];
    KSA(key, S);

    PRGA(S, plaintext, ciphertext);

    return 0;
}
\end{lstlisting}

The resulting LLVM IR function snippets in respective ISAs are shown below.

\newpage

\textbf{ARM}
\begin{lstlisting}[language=llvm]
define i32 @RC4(i32 %arg1, i32 %arg2, i32 %arg3) local_unnamed_addr {
dec_label_pc_107e8:
  %stack_var_-268 = alloca i32, align 4
  %0 = call i32 @KSA(i32 %arg1, i32* nonnull %stack_var_-268), 
     !insn.addr !59
  %1 = call i32 @PRGA(i32* nonnull %stack_var_-268, i32 %arg2, 
     i32 %arg3), !insn.addr !60
  ret i32 0, !insn.addr !61
}
\end{lstlisting}

\textbf{MIPS}
\begin{lstlisting}[language=llvm]
define i32 @RC4(i32 %arg1, i32 %arg2, i32 %arg3) local_unnamed_addr {
dec_label_pc_400b44:
  %stack_var_-268 = alloca i32, align 4
  %0 = call i32 @KSA(i32 %arg1, i32* nonnull %stack_var_-268)
    , !insn.addr !59
  %1 = call i32 @PRGA(i32* nonnull %stack_var_-268, i32 %arg2, 
    i32 %arg3), !insn.addr !60
  ret i32 0, !insn.addr !61
}
\end{lstlisting}

\textbf{x86}
\begin{lstlisting}[language=llvm]
define i32 @RC4(i32 %arg1, i32 %arg2, i32 %arg3) local_unnamed_addr {
dec_label_pc_1425:
  %eax.0.reg2mem = alloca i32, !insn.addr !94
  %0 = call i32 @__decompiler_undefined_function_0()
  %1 = call i16 @__decompiler_undefined_function_1()
  %stack_var_-274 = alloca i32, align 4
  %2 = call i32 @__asm_sti(i16 %1), !insn.addr !95
  %3 = call i32 @__x86.get_pc_thunk.ax(i32 %0), !insn.addr !96
  %4 = call i32 @__readgsdword(i32 20), !insn.addr !97
  %5 = call i32 @KSA(i32 %arg1, i32* nonnull %stack_var_-274), 
  !insn.addr !98
  %6 = call i32 @PRGA(i32* nonnull %stack_var_-274, i32 %arg2, 
    i32 %arg3), !insn.addr !99
  %7 = call i32 @__readgsdword(i32 20), !insn.addr !100
  %8 = icmp eq i32 %4, %7, !insn.addr !100
  store i32 0, i32* %eax.0.reg2mem, !insn.addr !101
  br i1 %8, label %dec_label_pc_14ac, label %dec_label_pc_14a7,
    !insn.addr !101

dec_label_pc_14a7:  ; preds = %dec_label_pc_1425
  %9 = call i32 @__stack_chk_fail_local(), !insn.addr !102
  store i32 %9, i32* %eax.0.reg2mem, !insn.addr !102
  br label %dec_label_pc_14ac, !insn.addr !102

dec_label_pc_14ac:  ; preds = %dec_label_pc_14a7, %dec_label_pc_1425
  %eax.0.reload = load i32, i32* %eax.0.reg2mem
  ret i32 %eax.0.reload, !insn.addr !103
}

\end{lstlisting}

It is obvious from above that code inserted by the compiler for individual ISA such as stack canary checking for x86 (\texttt{\_\_stack\_chk\_fail\_local}) are also captured in the lifting process, which made the lifted LLVM IR from x86 a bit more different than ARM and MIPS.

Then IR tokens after the transformations from each ISA are shown below:

\textbf{ARM}
\begin{lstlisting}[language=LLVM]
%stack_var = alloca align 4
%r = call @KSA( %r nonnull %stack_var
%r = call @PRGA(* nonnull %stack_var %r %r
ret 0,
\end{lstlisting}

\textbf{MIPS}
\begin{lstlisting}[language=LLVM]
%stack_var = alloca align 4
%r = call @KSA( %r nonnull %stack_var 
%r = call @PRGA(* nonnull %stack_var %r %r 
ret 0,
\end{lstlisting}

\textbf{x86}
\begin{lstlisting}[language=LLVM]
%r = alloca
%r = call @__decompiler_undefined_function_0()
%r = call @__decompiler_undefined_function_1()
%stack_var = alloca align 4 
%r = call @__asm_sti( %r 
%r = call @__x86.get_pc_thunk.ax( %r 
%r = call @__readgsdword( 20), 
%r = call @KSA( %r nonnull %stack_var 
%r = call @PRGA(* nonnull %stack_var %r %r 
%r = call @__readgsdword( 20), 
%r = icmp eq %r %r 
store 0, %r 
br i1 %r label %dec_label label %dec_label 
%r = call @__stack_chk_fail_local(), 
store %r %r 
br label %dec_label 
%r = load %r 
ret %r
\end{lstlisting}

As shown, after the transformation process, the LLVM IR tokens from ARM and MIPS are exactly the same. This is due to their original similarity in the first place, since they are both RISC (Reduced Instruction Set Computer) architectures. However the difference from x86 is quite large, as x86 is better utilized by the compiler (GCC) with stack protection mechanisms and is a CISC (Complex Instruction Set Computer) architecture. The extra compiler artefacts could be removed to improve cross-architecture similarity between x86 and the other RISC architectures, but is not done here to preserve inserted code following the principle of least surprise.

\subsection{Comparing Hashing Algorithms}

There are fundamental differences between how the four hashing algorithms are used to compare similarity of input data, therefore different ways of applying them to compare similarity of the transformed IR tokens.

MinHash\cite{broder1997resemblanceMinHash} turns data into $p$ hashes by turning data into shingles, and pseudo-randomly choosing $p$ shingles to hash by splitting the data using a fixed sized rolling window, where $p$ is the number of permutations (with a consistent seed so that the same input always produces the same selection).  The distance between two inputs are calculated using the Jaccard distance, as in, the size of intersection of the two sets of $p$ hashes. Two inputs can only be compared for similarity if the MinHash sets were generated with the same $p$. For example, if $p = 64$, $64$ numbers are generated for each input. If two inputs have $18$ numbers in common with their MinHash sets, then the similarity is $18 / 64 = 0.28125$. This value is between 0 and 1, and an arbitrary threshold can be set to decide if two inputs are similar or not.

SimHash\cite{charikar2002similaritySimHash} uses rounding algorithms to output a sketch for estimating the cosine similarity measure between inputs. Given two inputs, SimHash produces a $t$-bit vector (e.g. 64 bit number), where the cosine distance is 0 to $t$, from 0 being the two inputs are nearly the same to $t$ being extremely different. For this experiment, 64 bit SimHash values are used, which means distance is between 0 to 64 and the smaller the distance, the more similar two inputs are.

SSDEEP/CTPH\cite{2006ssdeep} is widely used in the industry for matching  similar malware. Adopted from the spamsum\cite{tridgell2002spamsum} algorithm, has two components: the rolling hash and the traditional hash (FNV). The rolling hash updates on each byte of the input, and at certain trigger points generates a traditional hash of the block of data, and appends the base64 encoded six Least Significant Bytes of the traditional hash to the digest. The trigger points are hit at each blocksize, which is calculated initially based on a minimum blocksize and the length of the input. SSDEEP processes the input twice; once using the length of the input to generate the initial blocksize, and once using half of it, generating two digests (one for ``chunk" and one for ``doublechunk"). Two digests are then compared using string edit distance, producing a re-scaled score of 0-100. Two inputs are considered similar if the matching score is bigger than 0. Note that two inputs can only be compared using SSDEEP if their blocksize is the same. This means inputs with large size differences will not match.

TLSH\cite{oliver2013tlsh} aims to trump the industry standard of SSDEEP by addressing some of its limitations: 1) the fact that it is very easy to ``trick" SSDEEP into generating completely different digests given targeted byte swaps in the input (while remaining largely similar) 2) it has a much lower detection ratio due to the use of a scaled similarity ratio that shows a match when score is bigger than 0. TLSH processes the input data using a sliding window of size 5 to populate buckets of triplets (e.g. for 5 bytes A,B,C,D,E, form buckets ABC, ABD, ABE, ACD ..), then select 6 out of 10 possible triplets to apply the Pearson hash\cite{pearson1990fast} on. The distance between two TLSH hashes are calculated using modular difference between the quartile ratio of similar items in buckets. The algorithm is not fully described here; the sake of brevity please refer to the TLSH paper for more details. The notable difference is that TLSH, much like SimHash outputs a distance metric instead of a similarity score, and the distance score is unbounded. A distance of 0 means that they two inputs are very similar, and anything more than that means they are more different (at which point a variable threshold can be applied, much like SimHash). The paper reports that false positives significantly increase past the distance of 100. The limitations of TLSH is that only data more than 50 bytes are eligible for hashing.

To compare the effectiveness of the four hashing algorithms, which all can have variable thresholds, ROC (Receiver Operating Characteristic) curves are used with a selected range of thresholds specific to each hashing algorithm (for unbounded distance scores such as TLSH, up to twice the range of the advised threshold is used, which is $100 * 2$).

\begin{table}[t]
\caption{\textbf{Comparison of hashing algorithms with score/distance thresholds (n=893116,Positives=3473)}}

\label{hashtable}
\hspace{-2cm}
\resizebox{18cm}{!}{

\begin{tabular}{@{}lll|lll|lll|lll@{}}
\toprule
MinHash & TPR   & FPR   & SimHash & TPR   & FPR   & SSDEEP & TPR   & FPR   & TLSH & TPR   & FPR   \\ \midrule
5    & 0.998 & 0.919 & 39      & 0.992 & 0.984 & 5      & 0.625 & 0.05  & 200  & 0.944 & 0.671 \\
10     & 0.993 & 0.826 & 37      & 0.971 & 0.946 & 10     & 0.625 & 0.05  & 190  & 0.926 & 0.609 \\
15   & 0.976 & 0.728 & 35      & 0.926 & 0.866 & 15     & 0.62  & 0.05  & 180  & 0.908 & 0.545 \\
20     & 0.962 & 0.624 & 33      & 0.872 & 0.737 & 20     & 0.605 & 0.049 & 170  & 0.879 & 0.478 \\
25    & 0.95  & 0.519 & 31      & 0.808 & 0.574 & 25     & 0.594 & 0.047 & 160  & 0.843 & 0.412 \\
30     & 0.926 & 0.379 & 29      & 0.725 & 0.403 & 30     & 0.584 & 0.043 & 150  & 0.804 & 0.347 \\
35    & 0.907 & 0.278 & 27      & 0.65  & 0.272 & 35     & 0.574 & 0.037 & 140  & 0.768 & 0.286 \\
40     & 0.886 & 0.188 & 25      & 0.615 & 0.195 & 40     & 0.56  & 0.031 & 130  & 0.72  & 0.231 \\
45    & 0.854 & 0.116 & 23      & 0.593 & 0.158 & 45     & 0.543 & 0.026 & 120  & 0.674 & 0.18  \\
50     & 0.791 & 0.066 & 21      & 0.579 & 0.14  & 50     & 0.508 & 0.02  & 110  & 0.621 & 0.137 \\
55    & 0.681 & 0.028 & 19      & 0.565 & 0.124 & 55     & 0.468 & 0.017 & 100  & 0.572 & 0.101 \\
60     & 0.604 & 0.017 & 17      & 0.553 & 0.102 & 60     & 0.434 & 0.014 & 90   & 0.523 & 0.071 \\
65    & 0.533 & 0.011 & 15      & 0.533 & 0.075 & 65     & 0.371 & 0.011 & 80   & 0.479 & 0.047 \\
70     & 0.463 & 0.008 & 13      & 0.516 & 0.049 & 70     & 0.319 & 0.008 & 70   & 0.423 & 0.029 \\
75    & 0.396 & 0.007 & 11      & 0.488 & 0.032 & 75     & 0.266 & 0.005 & 60   & 0.36  & 0.016 \\
80     & 0.321 & 0.004 & 9       & 0.462 & 0.023 & 80     & 0.236 & 0.003 & 50   & 0.322 & 0.008 \\
85    & 0.271 & 0.003 & 7       & 0.43  & 0.019 & 85     & 0.204 & 0.002 & 40   & 0.272 & 0.003 \\
90     & 0.23  & 0.002 & 5       & 0.356 & 0.013 & 90     & 0.186 & 0.002 & 30   & 0.224 & 0.001 \\
95    & 0.2   & 0.001 & 3       & 0.238 & 0.005 & 95     & 0.165 & 0.001 & 20   & 0.191 & 0.001 \\
100       & 0.169 & 0.001 & 1       & 0.174 & 0.001 & 100    & 0     & 0     & 10   & 0.161 & 0     \\ \bottomrule
\end{tabular}}
\end{table}

Table \ref{hashtable} shows the True Positive Rate and False Positives Rate of the different algorithms. Note that for SimHash and TLSH, the thresholds are distance based (smaller means more similar), while for MinHash and SSDEEP they are normalized scores from 0-100 percent (bigger means more similar). 

\begin{center}
\begin{figure}[ht]
\caption{ROC Curve} 
\hspace{+2.5cm}
\includegraphics[width=8cm]{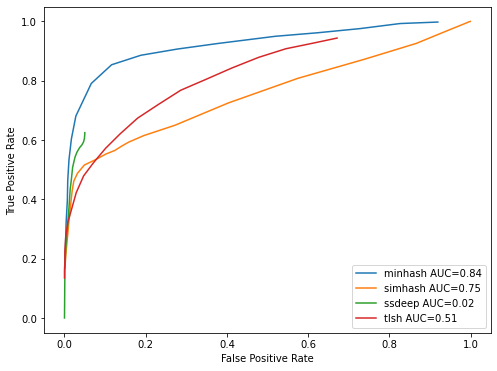}

\label{roc}
\end{figure}
\end{center}

Figure \ref{roc} plots the TPR (True Positive Rate) against the FPR (False Positive Rate) under varying thresholds using a ROC (Receiver Operating Characteristic) curve, which describes how well each algorithm performs under different thresholds (the bigger the Area Under the Curve, the better). From the graph it's evident that MinHash performs that best in this scenario, with TLSH in second and SSDEEP doing the worst (although SSDEEP had a very low False Positive Rate generally, its match rate was too low).

\subsection{Indexing and Searching}

As MinHash\cite{broder1997resemblanceMinHash} was the algorithm with the best ROC Curve and AUC (Area Under the Curve), it is chosen to implement a faster, more user friendly binary code search database. MinHash can be indexed to produce a hash table and enable O(1) lookups of each hash, since it produces a set of $p$ hashes from an input. After hashing the normalized LLVM IR tokens from a function and running them through MinHash, a set of $p$ integer hashes are produced. Each integer can be used as an index to store the function in a hash table, where the key is the integer and the value is a list of functions which contain the integer in its MinHash output hash.

So, given the MinHash output from a function's LLVM IR tokens, lookup of a function can be performed in best case $O(1)$ speed as $p$ is a constant hyperparameter  defined for MinHash. For a given binary with $n$ functions, lookup of similar functions with any matching hash in the entire database can be performed in $O(n)$, and if a threshold has to be applied to the number of hashes matched to determine Jaccard distance, we can store the count of each function matched in a hash table using the function name as the index, and perform at most $F$ hash table lookups to determine if the count of hashes matched is over a certain threshold for $F$ being the total number of functions in the database, making the worst case for getting similar functions in the database for a given binary $O(n * F * p)$.

The algorithm for indexing and lookup are described in Python pseudo-code below.

\begin{lstlisting}[language=python]
def index(func_llvmir):
    func_tokens = tokenize(func_llvmir) #the IR normalization process
    func_minhash = minhash(func_tokens) # produce minhash set
    for h in func_minhash:
      if db.get(h) == None:
        db[h] = []
      db[h].append(file_func)

def lookup(func_llvmir):
    func_tokens = tokenize(func_llvmir)
    func_minhash = minhash(func_tokens)
    matches = {}
    for h in func_minhash:
      if db.get(h) != None:
        matched = db.get(h)
        for func in matched:
          matches[func] += 1
\end{lstlisting}

The hyperparameter $p$, number of hashes in the output produced by MinHash can be tuned for a balance of accuracy and speed trade-offs (doubling $p$ means doubling the amount of hash lookups).

To find the best threshold, a small range of $p$ are used to test accuracy (in this case, due to the severe class imbalance in the dataset of having a lot more functions with different than same names, Balanced Accuracy (BA) is used to measure success: $\frac{(TPR + TNR)}{2}$

The $TPR$ (True Positive Rate) and $TNR$ (True Negative Rate) are calculated using a threshold of $0.5$. As shown in Table \ref{permu}, $p = 256$ is the optimal value.


\begin{table}[h]
\caption{Permutations}
\label{permu}
\begin{center}
\begin{tabular}{@{}llll@{}}
\toprule
\textbf{$p$} & \textbf{TPR} & \textbf{TNR} & \textbf{Balanced Accuracy} \\ \midrule
64         & 0.791        & 0.934        & 0.863             \\
128        & 0.817        & 0.929        & 0.873             \\
\textbf{256}        & 0.822        & 0.928       & \textbf{0.875}             \\
512        & 0.806        & 0.944        & 0.875             \\ \bottomrule
\end{tabular}
\end{center}
\end{table}

\subsection{Cross-Architecture Containment Detection}

From the experiments above it is clear that the LLVM IR + MinHash approach performs well to detect similar functions across architectures. However, can it detect code in smaller functions being embedded bigger ones? (Detecting binary code containment instead of just similarity/resemblance.) This is important and useful in aiding the reverse engineering process for several reasons:
\begin{enumerate}
    \item Code from a function could be copied and adapted, in part or in whole, into another bigger function
    \item Malware authors often insert dead code or no-operation instructions into start / end / middle of functions to avoid detection
    \item To patch a vulnerability, developers often add or remove code from an existing function for additional security, sometimes changing its size significantly.
\end{enumerate}

In theory, this should be detectable via MinHash, since the algorithm selects pseudo-random fixed sized pieces (shingles) from the input to hash and include in the resulting set \cite{broder1997resemblanceMinHash} in order to detect both resemblance and containment. The limitation to MinHash is that since random pieces are selected, there will always be data loss if the number of shingles are bigger than the hyperparameter $p$. This also means the bigger $p$ is, the better containment detection should be, and that smaller functions should have higher accuracy than larger ones.

To validate if this approach works for both cross-architecture similarity and containment, an experiment is done by modifying the RC4 code used in section \ref{irnormal} \cite{rc4gist} and combining the PRGA, KSA and RC4 wrapper function into one. The resulting combined function is listed below and compiled into Intel x86, ARM and MIPS architecture for comparison. 

\footnote{See the original code for the separate KSA and PRGA functions \cite{rc4gist}.}

\begin{lstlisting}[language=C]
int RC4(char *key, char *plaintext, unsigned char *ciphertext) {

    unsigned char S[N];

    int len = strlen(key);
    int j = 0;

    for(int i = 0; i < N; i++)
        S[i] = i;

    for(int i = 0; i < N; i++) {
        j = (j + S[i] + key[i % len]) % N;

        swap(&S[i], &S[j]);
    }

    int i = 0;
    j = 0;

    for(size_t n = 0, len = strlen(plaintext); n < len; n++) {
        i = (i + 1) % N;
        j = (j + S[i]) % N;

        swap(&S[i], &S[j]);
        int rnd = S[(S[i] + S[j]) % N];

        ciphertext[n] = rnd ^ plaintext[n];

    }

    // original wrapper calls commented out
    // KSA(key, S);
    // PRGA(S, plaintext, ciphertext);

    return 0;
}

\end{lstlisting}

To compare both cross-architecture resemblance and containment, the same process described in section \ref{approach} is applied to the original PRGA and KSA function, as well as the new combined RC4 function, and the MinHash outputs are compared across the different architectures they have been compiled in to produce a Jaccard similarity score between the contained and combined function in each architecture. To demonstrate the effects of changing the hyperparam $p$, the comparison has been done using multiple $p$ values. 

Function names could not be used as ground truth in x86 as the RetDec lifting process had a critical flaw that lead to the correct function body being split up from the correct name to one unnamed function immediately below it (in memory address). This threat to validity is discussed further in section \ref{threatsvalid}. Since this is a small dataset, the correct x86 functions were manually identified and compared to ensure results are sound.

\begin{table}[]
\caption{Cross-Architecture Containment and Resemblance, ($p$=64)}
\label{contain64}
\begin{tabular}{l|rrr}
\multicolumn{1}{c|}{} & \multicolumn{1}{c}{\textbf{Combined ARM}} & \multicolumn{1}{c}{\textbf{Combined MIPS}} & \multicolumn{1}{c}{\textbf{Combined x86}} \\ \hline
KSA ARM                            & \textbf{0.65625}                          & 0.53125                           & 0.421875                                  \\
KSA MIPS                           & 0.53125                          & \textbf{0.734375}                          & 0.484375                                  \\
KSA x86                            & 0.25                                      & 0.328125                                   & \textbf{0.796875}                         \\
PRGA ARM                           & \textbf{0.625}                            & 0.5                               & 0.390625                                  \\
PRGA MIPS                          & 0.546875                         & \textbf{0.78125}                           & 0.40625                                   \\
PRGA x86                           & 0.46875                                   & 0.515625                          & \textbf{0.578125}                        
\end{tabular}
\end{table}

\begin{table}[]
\caption{Cross-Architecture Containment and Resemblance, ($p$=128)}
\label{contain128}
\begin{tabular}{l|rrr}
\textbf{} & \multicolumn{1}{c}{\textbf{Combined ARM}} & \multicolumn{1}{c}{\textbf{Combined MIPS}} & \multicolumn{1}{c}{\textbf{Combined x86}} \\ \hline
KSA ARM   & \textbf{0.703125}                         & 0.59375                           & 0.4609375                                 \\
KSA MIPS  & 0.59375                          & \textbf{0.7578125}                         & 0.5                                       \\
KSA x86   & 0.3125                                    & 0.3359375                                  & \textbf{0.7890625}                        \\
PRGA ARM  & \textbf{0.6875}                           & 0.5703125                         & 0.4296875                                 \\
PRGA MIPS & 0.5703125                        & \textbf{0.8359375}                         & 0.3984375                                 \\
PRGA x86  & 0.5234375                        & 0.53125                           & \textbf{0.59375}                         
\end{tabular}
\end{table}

\begin{table}[]
\caption{Cross-Architecture Containment and Resemblance, ($p$=256)}
\begin{tabular}{l|rrr}
\textbf{} & \multicolumn{1}{c}{\textbf{Combined ARM}} & \multicolumn{1}{c}{\textbf{Combined MIPS}} & \multicolumn{1}{c}{\textbf{Combined x86}} \\ \hline
KSA ARM   & \textbf{0.69921875}                       & 0.57421875                        & 0.47265625                                \\
KSA MIPS  & 0.58984375                       & \textbf{0.765625}                          & 0.48828125                                \\
KSA x86   & 0.3359375                                 & 0.34375                                    & \textbf{0.77734375}                       \\
PRGA ARM  & \textbf{0.7109375}                        & 0.60546875                        & 0.41796875                                \\
PRGA MIPS & 0.59375                          & \textbf{0.84375}                           & 0.3984375                        \\
PRGA x86  & 0.5546875                        & 0.5703125                         & \textbf{0.5703125}                       
\end{tabular}
\end{table}

\begin{table}[]
\caption{Cross-Architecture Containment and Resemblance, ($p$=512)}
\begin{tabular}{l|rrr}
\textbf{} & \multicolumn{1}{c}{\textbf{Combined ARM}} & \multicolumn{1}{c}{\textbf{Combined MIPS}} & \multicolumn{1}{c}{\textbf{Combined x86}} \\ \hline
KSA ARM   & \textbf{0.716796875}                      & 0.599609375                       & 0.4453125                                 \\
KSA MIPS  & 0.59765625                       & \textbf{0.78515625}                        & 0.453125                                  \\
KSA x86   & 0.3046875                                 & 0.326171875                                & \textbf{0.771484375}                      \\
PRGA ARM  & \textbf{0.748046875}                      & 0.634765625                       & 0.40625                                   \\
PRGA MIPS & 0.6171875                        & \textbf{0.85546875}                        & 0.388671875                      \\
PRGA x86  & 0.533203125                      & 0.55859375                        & \textbf{0.5625}                          
\end{tabular}
\end{table}

\begin{table}[]
\caption{Cross-Architecture Containment and Resemblance, ($p$=1024)}
\label{contain1024}
\begin{tabular}{l|rrr}
\textbf{} & \multicolumn{1}{c}{\textbf{Combined ARM}} & \multicolumn{1}{c}{\textbf{Combined MIPS}} & \multicolumn{1}{c}{\textbf{Combined x86}} \\ \hline
KSA ARM   & \textbf{0.712890625}                      & 0.576171875                       & 0.435546875                               \\
KSA MIPS  & 0.5849609375                     & \textbf{0.7919921875}                      & 0.4248046875                              \\
KSA x86   & 0.314453125                               & 0.322265625                                & \textbf{0.7734375}                        \\
PRGA ARM  & \textbf{0.751953125}                      & 0.619140625                       & 0.388671875                               \\
PRGA MIPS & 0.611328125                      & \textbf{0.8603515625}                      & 0.3740234375                              \\
PRGA x86  & 0.529296875                      & 0.541015625                       & \textbf{0.556640625}                     
\end{tabular}
\end{table}

The same comparisons were ran across $p$ values 64,128,256,512 and 1024, shown from tables \ref{contain64} to \ref{contain1024}. The containment comparison scores across the same architecture are in bold, and those values show that same-architecture containment detection works. Cross-architecture containment detection was successful between MIPS and ARM (50 to 60 percent match), which are both RISC (Reduced Instruction Set Computer) architectures; but was only moderately successful (above 40 percent) when those two architectures are compared against x86, a CISC (Complex Instruction Set Computer) architecture, evidently shown by the average scores between the three architectures across increasing $p$ values on table \ref{avgcontain}. 

The \href{https://gist.github.com/rverton/a44fc8ca67ab9ec32089#file-rc4-c-L18}{KSA function} (line 18 in \cite{rc4gist}) is significantly smaller in size when compiled to ARM and MIPS versus in x86. This is because the Key Scheduling Algorithm is mainly an I/O bound function with many memory movement instructions in loops, which produces more instructions in a RISC architecture that does not have looped memory movement mnemonics unlike in x86. Since the x86 compiled assembly is smaller, and the instructions are different, the cross-architecture detection rates are very low for the KSA function as seen on tables \ref{contain64} to \ref{contain1024}. On the other hand, the \href{https://gist.github.com/rverton/a44fc8ca67ab9ec32089#file-rc4-c-L35}{PRGA function} mostly consists of mathematical operations (like modulus and XOR), which produces more similar tokens when lifted to LLVM IR, and thus have a higher similarity between RISC and CISC architectures. These findings show that BCD's approach captures underlying design differences between instruction set architectures (ISAs), which can affect cross-architecture comparison results between classes of ISAs.

It is also worth noting that increasing the $p$ value does not mean the detection rate will always increase; since MinHash selects $p$ random shingles to include, having more shingles provide more accurate similarity scores, but not always consistently due to the random selection (on table \ref{avgcontain}, the average score increased significantly after $p > 64$, but then fluctuates.

\begin{table}[bt]
\caption{Average containment detection scores}
\label{avgcontain}
\begin{center}
\begin{tabular}{r|rrr}
\multicolumn{1}{l|}{\textbf{p}} & \multicolumn{1}{l}{\textbf{arm vs mips}} & \multicolumn{1}{l}{\textbf{arm vs x86}} & \multicolumn{1}{l}{\textbf{mips vs x86}} \\ \hline
64                                         & 0.52734375                                   & 0.3828125                                   & 0.43359375                                   \\
128                                        & 0.58203125                                   & 0.431640625                                 & 0.44140625                                   \\
256                                        & 0.5908203125                                 & 0.4453125                                   & 0.4501953125                                 \\
512                                        & 0.6123046875                                 & 0.4223632813                                & 0.431640625                                  \\
1024                                       & 0.5979003906                                 & 0.4169921875                                & 0.4155273438                                
\end{tabular}
\end{center}
\end{table}

\subsection{Usability}

As the goal of BCD is to aid reverse engineers in their initial stages of the RE process, the availability and the usability of code is key. To the loss of the reverse engineering field, many prior works in Binary Code Similarity did not open source their code (only 16 out of 70 evaluated approaches were open sourced according to the 2021 survey\cite{haq2021survey}).

To make BCD as usable as possible, two different interfaces to use BCD are made available: a command line program that takes arguments (as many reverse engineers like using the terminal, and command line programs can be integrated into automation tools such as plugins and CI/CD pipelines), as well as a web interface so that BCD can be ran over the network as a shared service for analysts and less advanced users. The command line program is  \texttt{bcd.py}, and the Flask web application is \texttt{server.py}, and they are both open sourced on GitHub \href{https://github.com/h4sh5/bcddb}{here}.

To index functions in a binary using the command line program, simply use the index mode and point the program either at the path of the binary or the lifted LLVM IR file (if the binary is passed, it will automatically be lifted with RetDec\cite{retdecGH}:

\texttt{./bcd.py -i /path/to/binary}

To search all functions in a target binary, simply do the same but using \texttt{-s} (by default, the program uses search mode):

\texttt{./bcd.py -s /path/to/binary}

The web interface allows users to upload a binary and perform index or searching. For each function in the target binary, the functions that matched it are ranked based on the similarity score.

\subsection{Usage Recommendations}

BCD performs many-to-many function comparisons under the hood; as in, for each function in an input binary, it will compare the functions against all known functions in the database using MinHash. This means that the more functions are in a database, the longer comparisons will take. Therefore we recommend building different databases containing different types of functions to compare against, where each function has its original symbol name or is labelled. For example:
\begin{itemize}
    \item A database for known Linux library calls (index libc with symbols)
    \item A database for known cryptographic functions commonly used in malware
    \item A database with known malware samples
    \item A database with known functions vulnerable to a certain type of bug
    
\end{itemize}

Building separate databases for each use-case can not only reduce search time, but also make BCD act as a function tagging / multi-class classifier system. 

Another good use case is to compile an open-source project (with different compilers, optimization levels and architectures), then to search unknown functions against it to detect open-source functions in the binary; that way, function matches with high scores can effectively be ``decompiled". It can also be useful for code-clone detection purposes, such as detecting violation against open source licensing.

\section{Evaluation}

\subsection{Efficiency}

Efficiency testing was done on the hash table implementation of BCD (\texttt{bcd.py}) instead of the experiment script (which used a very slow one pair at a time comparison approach to compare functions). 

The dataset used is coreutils 8.32\cite{coreutils8.32}, compiled across four different architectures (x64, aarch64, mips and powerpc), totally to 441 binaries and 81668 lifted functions. Out of those 441 binaries, 436 was successfully lifted and the rest had errors.

Time taken to lift each executable to LLVM IR varies as BCD uses on the efficiency of RetDec, but is generally quite fast ($<$ 1 minute), for example, lifting \texttt{/bin/ls} took 10 seconds and \texttt{/bin/sh} 7 seconds in x64 architecture.

Indexing the entire dataset already lifted to LLVM IR and constructing the hash table took 212.68 seconds. That means on average, indexing each binary's LLVM IR code into the database takes 0.48 seconds and indexing each function takes 0.0026 seconds (2.6 milliseconds). The data set produced 1363199 keys (each one a number in the MinHash set) in total (using $p=256$).

Searching functions in the database was tested by taking binaries to search against the database. The database hash table object is saved as a Python pickle file for speed of saving and loading. The database in total was 132MB in size, and took 15 seconds to load into memory. An example binary used was \texttt{/usr/bin/test} on a x64 Linux system, which was 55KB in size, had 227 lifted LLVM IR functions, and took 128 seconds to perform many-to-many comparisons (227 functions in the test binary against 81668 functions in the database). The same test done on \texttt{/usr/bin/gcc}, with 2616 lifted functions, took 1634.37 seconds. 

If we divide time by the number of total function pairs compared (which would be $227\times81668=18538636$ and $2616\times81668 = 213643488$ respectively), each function pair comparison only takes 0.006-0.007 milliseconds. This shows that BCD is scalable when it comes to one-to-many binary comparisons and many-to-many binary function searches. 

\subsection{Accuracy}

Accuracy testing was done using the same dataset, by extracting 10000 random functions that had symbols (not unnamed functions, which would be labeled \texttt{@function\_xxxx} by RetDec. Each function selected would be compared against the same number of same name functions and the different name functions to ensure classes are balanced. At threshold $t=0.5$ and permutations $p=256$, BCD achieved 0.808 accuracy, 0.910 precision and 0.685 recall ($F_1=0.782$, $F_2=0.721$). 

\section{Threats to Validity}
\label{threatsvalid}

One of the advantages of doing binary code similarity over source code similarity is that ``What You See Is Not What You eXecute" (WYSINWYX)\cite{wysinwyxWhatYouSee} - the compilation process is complicated and may add many additional code into the output binary. By lifting the compiled code into an IR that captures extra compiler outputs, these compilation ``side-effects" can be detected. However, due to time constraints, BCD has not been evaluated finding similar code against cross-optimization levels or cross-compiler differences.

The tool used to lift binary code to LLVM IR, RetDec, has flaws that can affect the results of BCD significantly. For example, the GCC compiler inserts \texttt{endbr} or \texttt{endbr64} instructions in Intel 32/64 bit architecture as a control flow integrity feature; but at times a bug in RetDec causes this to cut a decompiled function short, causing the function body in the decompiled code to be incomplete. Given that BCD uses function names as ground truth for accuracy measures, this can skew the results. However, as all prior work in this field, BCD relies on external tools for disassembling and lifting binary code, therefore propagating any mistakes the tools make into the results. It is worth noting that the general framework of BCD, since it uses string manipulation and fuzzy hashing of lifted IR code, can work with any other intermediate representation, source code as well as assembly code to perform comparisons.

\section{Conclusion and Future Work}
In this paper, a framework for detecting binary code similarity across different CPU architecture was designed, implemented and released as usable code. We compared multiple fuzzy hashing algorithms for accuracy to be used in this framework, and built an efficient approach to search similar (including embedded/contained) binary code in a large database of cross-architecture binary functions. As BCD aims to aid reverse engineers, usability considerations in the reverse engineering process are discussed along with use case suggestions.

In the future, cross-optimization and cross-compiler evaluations can be performed using BCD, and improving the framework to increase effectiveness of results especially by further fine tuning the IR Normalization process. Usability and design considerations around specific use cases for the binary code similarity matching should be further researched, as many prior works in this field did not publish their code as usable tooling nor consider the use cases in a reverse engineer centered manner. 

\bibliographystyle{ieeetr}
\bibliography{refs}

\end{document}